\documentclass{article}

\usepackage{microtype}
\usepackage{graphicx}
\usepackage{subcaption}
\usepackage{booktabs}
\usepackage{float}
\usepackage{hyperref}

\usepackage[preprint]{icml2026}
\usepackage{amsmath}
\usepackage{amssymb}
\usepackage{mathtools}
\usepackage{amsthm}

\usepackage[capitalize,noabbrev]{cleveref}

\theoremstyle{plain}

\theoremstyle{definition}

\theoremstyle{remark}

\icmltitlerunning{Failure Modes of Deep Multi-Agent RL in Asynchronous Pricing}

\begin{document}

\twocolumn[
  \icmltitle{Failure Modes of Deep Multi-Agent RL in Asynchronous Pricing:\\
             Reproducible Triggers, Trace Diagnostics, and a Partial Fix}

  \icmlsetsymbol{equal}{*}
  \begin{icmlauthorlist}
    \icmlauthor{Shree Murthy}{do,equal}
    \icmlauthor{Rohan Pandey}{do,equal}
  \end{icmlauthorlist}

  \icmlaffiliation{do}{DigitalOcean, USA}

  \icmlcorrespondingauthor{Shree Murthy}{smurthy@digitalocean.com}
  \icmlcorrespondingauthor{Rohan Pandey}{rpandey@digitalocean.com}

  \icmlkeywords{multi-agent reinforcement learning, algorithmic collusion,
                continuous-time MDPs, semi-MDPs, DDPG, failure modes,
                trace diagnostics}

  \vskip 0.3in
]

\printAffiliationsAndNotice{\icmlEqualContribution}

\begin{abstract}
We study two reproducible failure modes of deep multi-agent reinforcement
learning in continuous-time pricing markets: (i) tacit cartel formation
between competing DDPG agents, and (ii) actor--critic instability at high
event rates. We instantiate both inside a single CT-MARL benchmark
(Poisson-clocked price updates, observation latency $\delta$, interior-optimum
logit demand), show that synchronous DDPG agents reliably trigger Failure
Mode 1 with collusion index $\Delta = 0.69 \pm 0.11$, and quantify a partial
microstructure fix: asynchrony alone cuts collusion by 48\% and adding
latency drives it to a minimum of $\Delta = 0.28$. The fix has clearly
documented costs: it is partial ($\Delta$ remains supra-Bertrand), it is
non-monotone in $\delta$, and it does not survive Failure Mode 2, which
emerges as DDPG critic divergence at $\lambda = 5$ and corrupts the
phase-diagram cell at $(\lambda{=}5, \delta{=}1)$. We accompany the
scalar collusion index with trajectory-level trace diagnostics that expose
the within-episode signalling collapse and the post-shock non-recovery.
\end{abstract}

\section{Introduction}\label{sec:intro}

Multi-agent reinforcement learning (MARL) systems are increasingly being
deployed in mixed-motive economic settings where the boundary between
``competitive equilibrium'' and ``coordinated harm'' is not a property
of the algorithm, but an emergent property of the environment that the
agents learn in. \citet{calvano2020} showed that tabular Q-learning
agents charging prices in a synchronous Bertrand oligopoly converge to
supra-competitive prices sustained by reward-and-punishment strategies
without explicit communication. Subsequent work has extended this to
deep RL methods \citep{hettich2024,deng2024}, alternative agent
architectures including LLMs \citep{fish2024llm}, and macroeconomic
shocks \citep{veres2025inflation}. Surveys identify ``asynchronous
timing, communication latency, and event-driven pricing'' as understudied
robustness conditions \citep{bichler2025review,deng2024}.

We adopt the failure-mode lens of recent agentic-AI evaluation work: a
failure mode is a reproducible behaviour, elicited by a small number of
named environment and agent factors, with a verifiable trace and a
partially understood cost-aware mitigation. The substantive
contribution of this paper is to characterise two such failure modes in
CT-MARL pricing:

\textbf{Failure Mode 1 (FM1): tacit cartel formation.} Two DDPG agents
trained in a continuous-time Bertrand duopoly converge to prices well
above the Bertrand--Nash equilibrium. The mode is robust across seeds
and is detectable from a single scalar (the collusion index $\Delta$)
plus a visible plateau in the per-episode price trace.

\textbf{Failure Mode 2 (FM2): critic instability at high event rate.}
At $\lambda = 5$ Poisson-clocked events per agent per unit time, the
DDPG critic loses stability before the buffer accumulates enough
trajectory diversity, and the policy diverges (mean prices spiralling
above the monopoly level on a subset of seeds). The mode is
reproducible on the corresponding $(\lambda, \delta)$ phase-diagram
cell.

We quantify both failure modes on a single benchmark, supply
trace-level diagnostics for each, and study a candidate mitigation for
FM1 (continuous-time market microstructure: Poisson asynchrony plus
observation latency) under which collusion drops by 48--59\%. The fix
is partial: $\Delta$ remains substantially above Bertrand--Nash, the
effect is non-monotone in $\delta$, and it does not address FM2. We
argue this is the kind of result the failure-mode framing is designed
to surface: a precisely characterised, reproducibly triggered
behaviour; a verifiable trace; a fix with bounded but documented effect.

\paragraph{Contributions.}
\begin{itemize}\itemsep0pt
\item A self-contained, reproducible CT-MARL pricing benchmark with
  interior-optimum logit demand, Poisson-clocked agents, observation
  latency, and an optional transient demand shock.
\item A semi-MDP DDPG agent that uses the continuous-time discount
  $\gamma_\tau = e^{-\rho \tau}$ over the agent's own sojourn time
  $\tau$, extending standard DDPG to the semi-MDP setting with a
  continuous pricing action space.
\item A 16-cell $(\lambda, \delta)$ phase diagram of $\Delta$, plus
  eight headline conditions with five seeds each, isolating FM1 and FM2.
\item Trajectory-level trace diagnostics complementing the scalar
  collusion index, including a stress-condition trajectory that
  exposes post-shock non-recovery.
\item A partial mitigation of FM1 with cost-of-fix accounting.
\end{itemize}

\section{Related Work}\label{sec:rw}

The foundational result is~\citet{calvano2020}: tabular $Q$-learning
agents in a synchronous Bertrand--Edgeworth oligopoly learn
supra-competitive strategies. \citet{klein2021} shows the result is
robust to sequential moves. \citet{hettich2024} and \citet{deng2024}
extend to deep RL methods (PPO, DQN, DDPG) and find generally weaker
but still super-competitive collusion. \citet{paudel2024} apply the
same setup to EV-charging pricing. \citet{veres2025inflation} show
that inflation shocks reshape but do not eliminate collusion;
\citet{fish2024llm} replicate the phenomenon with LLM agents.

The methodological tools we use sit at the intersection of CT-MARL
and asynchronous decision-making. \citet{wang2026ctmarl} introduce
continuous-time value iteration in a general MARL setting.
\citet{du2020smdpneuralode} formalise sojourn-time decisions as
semi-MDPs solved with neural ODEs. \citet{xiao2025async} construct
macro-action MARL that re-decides on event triggers in cooperative
robotics. \citet{sutton1999options} provide the classical bridge
between MDPs and semi-MDPs that we adopt. To our knowledge, none of
these threads has been applied to the algorithmic-collusion question,
and none has documented FM2-style critic instability in the
high-event-rate regime.

\section{Background}\label{sec:background}

\paragraph{Continuous-time duopoly.}
Two firms $i \in \{1,2\}$ post prices
$p_i \in [\underline{p}, \overline{p}]$ in real time
$t \in [0, T_{\max}]$. The instantaneous logit demand for firm $i$ at
$(p_1, p_2)$ is
\begin{equation}
q_i(p_i, p_{-i}) = \frac{e^{(1-p_i)/\mu}}{1 + e^{(1-p_i)/\mu} + e^{(1-p_{-i})/\mu}},
\end{equation}
with marginal cost $c=0$. Profit per unit time is
$\pi_i = (p_i - c) q_i$. We set $\mu = 0.25$, which yields a
symmetric Bertrand--Nash price $p_{\text{BN}} \approx 0.473$ and a
symmetric joint-monopoly price $p_M \approx 0.925$, both strictly
interior in the action space $[0, 2]$. This rules out the
``hit-the-ceiling'' artefact in which collusion is bounded by an
arbitrary action-space upper bound.

\paragraph{Poisson clocks and latency.}
Each firm $i$ has an independent Poisson event clock with rate
$\lambda_i$. Between events its price is held constant. Each firm
observes the rival's price with delay $\delta \ge 0$:
$\hat p_{-i}(t) = p_{-i}(t - \delta)$. A demand shock multiplies
$q_i$ by $\sigma_{\text{shock}} \in (0, 1)$ during an interval
$[t_s, t_s + \Delta_s]$.

\paragraph{Semi-MDP formulation.}
The decision process for firm $i$ is a semi-MDP
\citep{sutton1999options}. At each event $k$, the agent observes
$s_i^k$, chooses action $a_i^k = p_i$, and receives the integrated
reward $R_i^k = \int_{t_i^k}^{t_i^{k+1}} \pi_i(t)\,dt$ over its
sojourn $\tau = t_i^{k+1} - t_i^k$. The continuous-time discount is
$\gamma_\tau = e^{-\rho \tau}$.

\paragraph{Failure-mode metric: collusion index.}
For policies inducing average prices $\bar p_1, \bar p_2$ over the
evaluation window, with $\bar p = (\bar p_1 + \bar p_2)/2$, define
\begin{equation}
\Delta = \frac{\bar p - p_{\text{BN}}}{p_M - p_{\text{BN}}}.
\end{equation}
$\Delta = 0$ corresponds to the Bertrand--Nash equilibrium and
$\Delta = 1$ to perfect collusion at the joint monopoly price.\footnote{%
We define $\Delta$ in \emph{price space}. \citet{calvano2020} use a
profit-space index
$\Delta_\pi = (\bar\pi - \pi_{\text{BN}})/(\pi_M - \pi_{\text{BN}})$.
Under symmetric play the two are ordinally equivalent (profit is
monotone increasing in price on $[p_{\text{BN}}, p_M]$) but
numerically distinct: for this demand specification, $\Delta_\pi$
runs approximately $0.15$--$0.20$ higher than $\Delta$ across all
conditions.}
We treat any $\Delta \gtrsim 0.5$ in expectation across seeds as a
positive trigger of Failure Mode 1.

\section{Method}\label{sec:method}

\paragraph{Agent architecture.}
Each agent is a semi-MDP Deep Deterministic Policy Gradient (DDPG)
agent \citep{lillicrap2016ddpg} with a continuous price action
$a \in [0, 2]$. The \emph{actor}
$\pi_\theta : \mathcal{S} \to [p_{\min}, p_{\max}]$ is a two-layer MLP
(64 units, ReLU) with a $\tanh$ output scaled to
$[p_{\min}, p_{\max}]$. The \emph{critic} $Q_\omega(s, a)$ is a
two-layer MLP (64 units, ReLU) that takes the concatenation
$(s, a)$. Both networks have separate target networks
$(\theta', \omega')$ soft-updated at rate $\tau_{\text{soft}} = 0.005$.

\paragraph{Semi-MDP update.}
At each pricing event the agent stores the transition
$(s, a, R, s', \tau)$ in a replay buffer of capacity $50{,}000$.
After each event, a mini-batch of 128 transitions is sampled every
four steps and the critic is updated by minimising
\begin{align}
\mathcal{L}(\omega) &= \mathbb{E}\bigl[(Q_\omega(s, a) - y)^2\bigr], \\
y &= R + e^{-\rho \tau}\,Q_{\omega'}(s', \pi_{\theta'}(s')),
\end{align}
where the discount $\gamma_\tau = e^{-\rho\tau}$ is sojourn-dependent,
implementing the semi-MDP discounting from \cref{sec:background}. The
actor is updated by deterministic policy gradient:
\begin{equation}
\nabla_\theta J \approx \mathbb{E}\bigl[-\nabla_a Q_\omega(s,a)\big|_{a=\pi_\theta(s)}
\cdot \nabla_\theta \pi_\theta(s)\bigr].
\end{equation}
Exploration uses Gaussian noise $\mathcal{N}(0, \sigma^2)$ added to
the deterministic action, with $\sigma$ decaying from $0.20$ to
$0.02$ at rate $0.9998$ per step. Gradient norms are clipped to
$1.0$; we use Adam optimisers with learning rates $10^{-4}$ (actor)
and $10^{-3}$ (critic), and $\rho = 0.05$. The integrated reward $R$
is computed by 5-point trapezoidal integration of $\pi_i$ over the
agent's sojourn.

\begin{algorithm}[h]
\caption{CT-MARL-DDPG (one episode, one agent shown).}
\label{alg:ctmarl}
\begin{algorithmic}
\STATE {\bfseries Require:} env $(T_{\max}, \lambda, \delta)$;
DDPG params $(\theta, \omega, \theta', \omega')$; buffer $\mathcal{B}$
\STATE Reset env; $s_{\text{prev}}[i] \leftarrow$ \texttt{None}
\WHILE{$t < T_{\max}$}
    \STATE $(i, s, R, \texttt{done}) \leftarrow \texttt{env.step\_to\_next\_event}()$
    \IF{$s_{\text{prev}}[i] \neq \texttt{None}$}
        \STATE $\tau \leftarrow t - t_{\text{prev}}[i]$
        \STATE $\mathcal{B}.\texttt{push}(s_{\text{prev}}[i], a_{\text{prev}}[i], R, s, \tau)$
        \IF{$|\mathcal{B}| \geq B$ and step $\equiv 0 \!\!\pmod{4}$}
            \STATE Sample mini-batch $\{(s_j, a_j, R_j, s'_j, \tau_j)\}$
            \STATE $\gamma_j \leftarrow e^{-\rho \tau_j}$
            \STATE $y_j \leftarrow R_j + \gamma_j Q_{\omega'}(s'_j, \pi_{\theta'}(s'_j))$
            \STATE Critic step: minimise $\sum_j (Q_\omega(s_j, a_j) - y_j)^2$
            \STATE Actor step: minimise $-\sum_j Q_\omega(s_j, \pi_\theta(s_j))$
            \STATE Soft-update $\theta', \omega'$
        \ENDIF
    \ENDIF
    \IF{not \texttt{done}}
        \STATE $a \leftarrow \pi_\theta(s) + \mathcal{N}(0, \sigma^2)$,\; clipped
        \STATE $\texttt{env.apply\_action}(i, a)$
        \STATE $s_{\text{prev}}[i] \leftarrow s$;\; $a_{\text{prev}}[i] \leftarrow a$;\; $t_{\text{prev}}[i] \leftarrow t$
    \ENDIF
\ENDWHILE
\end{algorithmic}
\end{algorithm}

\paragraph{Trace diagnostic.}
We report per-episode price trajectories alongside the scalar
$\Delta$. Trace plots reveal mechanism (when in the episode the
collusive plateau forms, whether prices recover after a shock) that
final-window averages necessarily collapse. We use them in
\cref{sec:exp} to characterise both FM1 and the stress-condition
trajectory.

\section{Experiments}\label{sec:exp}

\paragraph{Conditions.} Eight headline conditions plus a $4 \times 4$
$(\lambda, \delta)$ phase-diagram sweep:
S0 (synchronous, $\Delta t = 0.5$); A0--A3 (async with
$\delta \in \{0, 0.5, 1.0, 2.0\}$, $\lambda = 1$); AR (slow,
$\lambda = 0.5$, $\delta = 1$); AF (fast, $\lambda = 5$, $\delta = 1$);
ST (async, $\lambda = 1$, $\delta = 1$, demand shock $0.4\times$ over
$t \in [25, 35]$). All conditions: $T_{\max} = 60$, 300 episodes per
seed, five seeds per main condition, three seeds per phase-diagram
cell. We report $\bar p_1, \bar p_2$ averaged over the last 40
episodes.

\paragraph{Baselines.} The static Bertrand--Nash and Monopoly price
constants supply the $\Delta$ scale; we additionally report the
synchronous-only baseline S0 to anchor the discrete-time, Calvano-style
result.

\subsection{FM1 Trigger and a Partial Microstructure Fix}
\label{sec:fm1}

\Cref{tab:main} summarises the seven stable main conditions (AF
discussed separately in \cref{sec:fm2}). The synchronous baseline S0
achieves $\Delta = 0.69 \pm 0.11$, confirming that DDPG agents learn
strong tacit collusion in the Calvano-style synchronous environment:
Failure Mode 1 is reliably triggered, providing the upper anchor for
the $\Delta$ scale.

Switching to Poisson-clocked asynchrony alone (A0, $\delta = 0$)
reduces collusion to $\Delta = 0.36 \pm 0.08$, a 48\% reduction.
Paired across seeds, $\Delta_{S0} - \Delta_{A0} = 0.33$ (95\%~CI
$[0.14, 0.53]$, $t(4) = 4.68$, $p < 0.01$). Increasing observation
latency from $\delta = 0$ to $\delta = 1.0$ pushes collusion to its
minimum: $\Delta_{A2} = 0.28 \pm 0.09$. The cumulative S0 to A2 effect
is a 59\% reduction ($\Delta_{S0} - \Delta_{A2} = 0.41$, 95\%~CI
$[0.19, 0.64]$, $t(4) = 5.07$, $p < 0.01$). \Cref{tab:stats} reports
all six paired contrasts; every one is significant at $p < 0.01$. At
very high latency ($\delta = 2.0$, condition A3), $\Delta$ rebounds
slightly to $0.30$, suggesting non-monotonicity at extreme delay that
warrants further study (\cref{sec:disc}).

\begin{table}[h]
\caption{Headline results across 5 seeds per condition (AF excluded;
see \cref{sec:fm2}). $\Delta = 0$ denotes Bertrand--Nash, $\Delta = 1$
denotes joint monopoly. The S0$\to$A0 (48\%) and S0$\to$A2 (59\%)
drops are both significant at $p < 0.01$ on a paired $t$-test.}
\label{tab:main}
\vskip 0.10in
\begin{center}\small
\begin{tabular}{lrrr}
\toprule
Condition & $\bar p$ & $\Delta$ & Reward \\
\midrule
S0 (sync)                              & 0.785 & $0.69 \pm 0.11$ & 18.87 \\
A0 ($\lambda{=}1$, $\delta{=}0$)       & 0.635 & $0.36 \pm 0.08$ & 16.50 \\
A1 ($\lambda{=}1$, $\delta{=}0.5$)     & 0.619 & $0.32 \pm 0.09$ & 15.97 \\
A2 ($\lambda{=}1$, $\delta{=}1.0$)     & 0.599 & $0.28 \pm 0.09$ & 15.74 \\
A3 ($\lambda{=}1$, $\delta{=}2.0$)     & 0.609 & $0.30 \pm 0.10$ & 15.93 \\
AR (slow, $\lambda{=}0.5$)             & 0.625 & $0.34 \pm 0.09$ & 16.25 \\
ST (shock)                             & 0.625 & $0.34 \pm 0.07$ & 14.56 \\
\bottomrule
\end{tabular}
\end{center}
\vskip -0.10in
\end{table}

\begin{table}[h]
\caption{Paired $t$-tests, S0 vs.\ each CT condition ($n=5$ seeds).
Positive values indicate that S0 achieves higher collusion.}
\label{tab:stats}
\vskip 0.10in
\begin{center}\small
\begin{tabular}{lrrr}
\toprule
Comparison & $\Delta_{S0}{-}\Delta_{\text{cond}}$ & $t(4)$ & $p$ \\
\midrule
S0 vs.\ A0 ($\delta{=}0$)   & $0.332$ & $4.68$ & $0.009$ \\
S0 vs.\ A1 ($\delta{=}0.5$) & $0.369$ & $6.15$ & $0.004$ \\
S0 vs.\ A2 ($\delta{=}1.0$) & $0.412$ & $5.07$ & $0.007$ \\
S0 vs.\ A3 ($\delta{=}2.0$) & $0.390$ & $5.07$ & $0.007$ \\
S0 vs.\ AR ($\lambda{=}0.5$) & $0.356$ & $4.73$ & $0.009$ \\
S0 vs.\ ST (shock)           & $0.354$ & $4.77$ & $0.009$ \\
\bottomrule
\end{tabular}
\end{center}
\vskip -0.10in
\end{table}

\subsection{FM1 Trace Diagnostic}\label{sec:trace}

\begin{figure*}[h]
  \centering
  \begin{subfigure}[t]{0.48\linewidth}
    \centering
    \includegraphics[width=\linewidth]{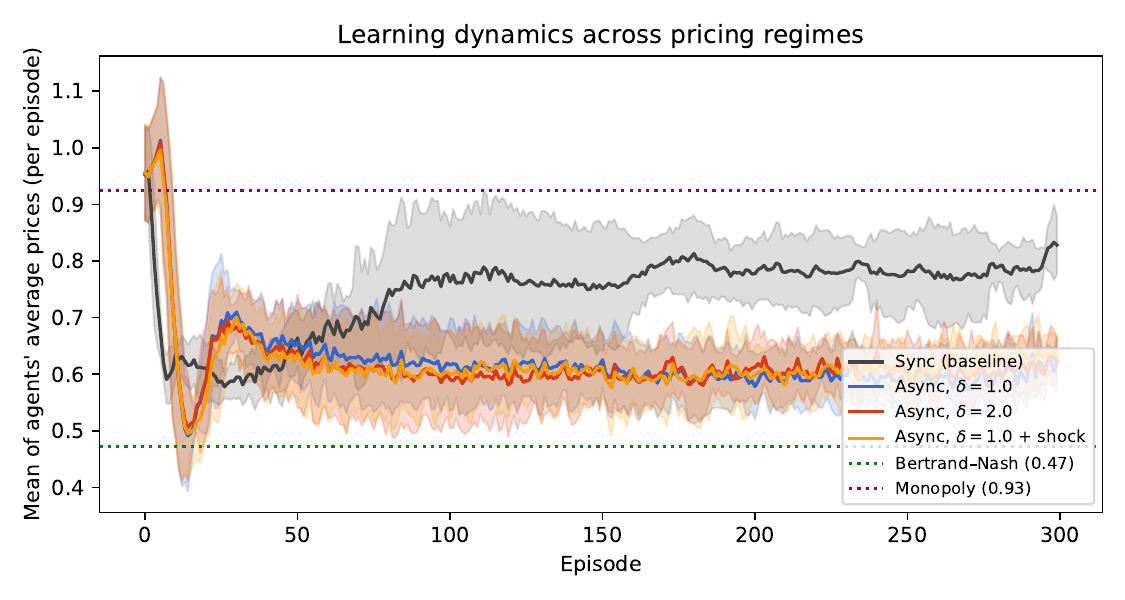}
    \caption{\textbf{Learning dynamics (FM1 trace).} Mean price per
      episode ($\pm 1$~SD across seeds). S0 climbs to the monopoly
      attractor; CT conditions plateau lower.}
    \label{fig:lc}
  \end{subfigure}
  \hfill
  \begin{subfigure}[t]{0.48\linewidth}
    \centering
    \includegraphics[width=\linewidth]{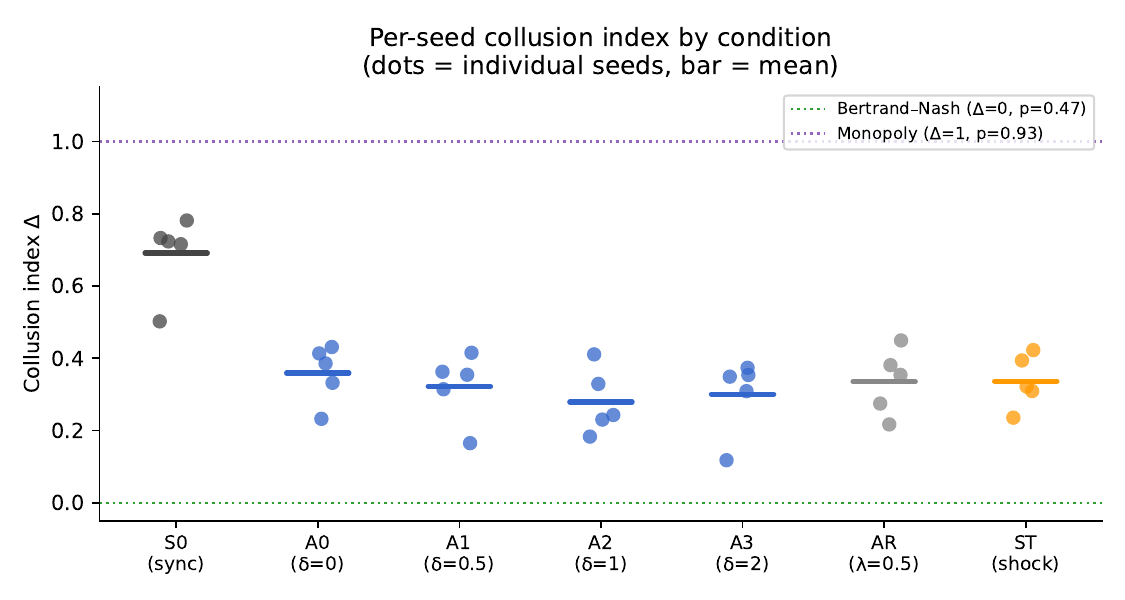}
    \caption{\textbf{Per-seed $\Delta$ by condition.} Dots = seeds,
      bar = mean. S0 ($\Delta = 0.69$) is highest; all CT conditions
      substantially lower.}
    \label{fig:bars}
  \end{subfigure}

  \vspace{1.0em}

  \begin{subfigure}[t]{0.45\linewidth}
    \centering
    \includegraphics[width=\linewidth]{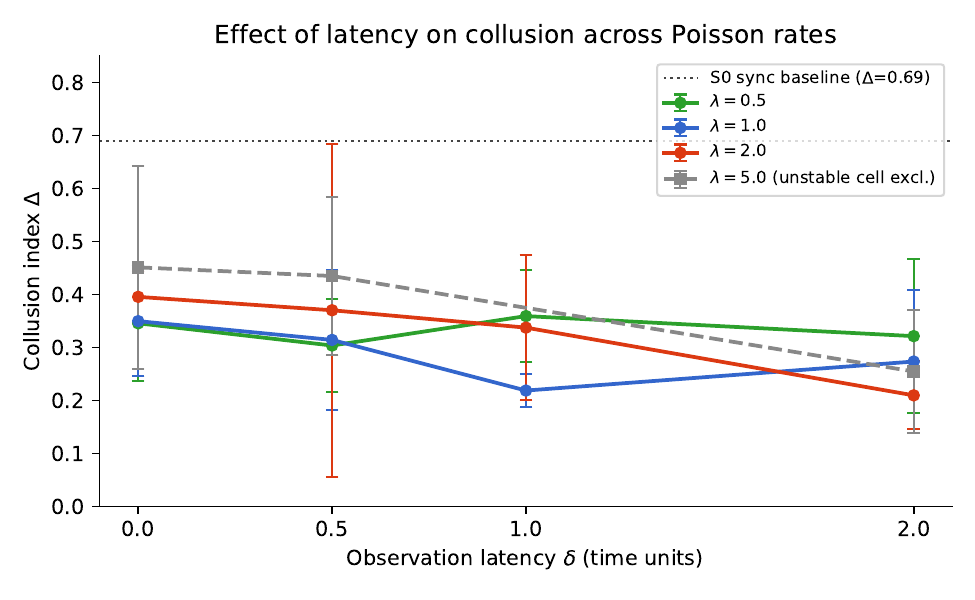}
    \caption{\textbf{Phase diagram.} $\Delta$ generally falls with
      latency; $\lambda = 5, \delta = 1$ is excluded as FM2.}
    \label{fig:phase}
  \end{subfigure}
  \hfill
  \begin{subfigure}[t]{0.51\linewidth}
    \centering
    \includegraphics[width=\linewidth]{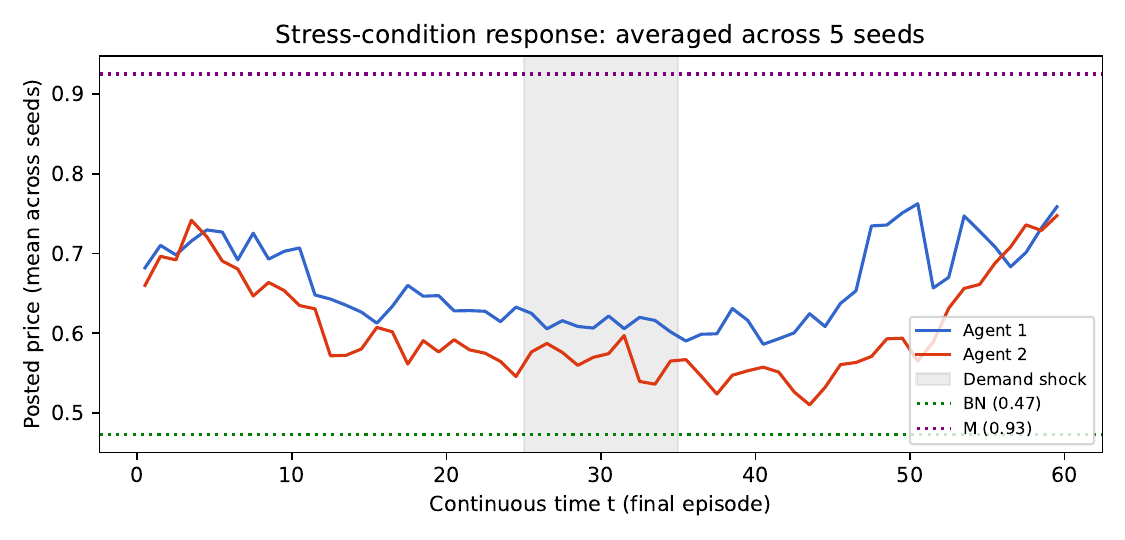}
    \caption{\textbf{Stress trace.} Prices fall during the shock
      window $t \in [25, 35]$ and do not recover.}
    \label{fig:shock}
  \end{subfigure}

  \caption{\textbf{Empirical results.} Trace-level diagnostics (a, d)
    complement the scalar collusion index (b) and the phase diagram
    (c). See \cref{sec:exp}.}
  \label{fig:results}
\end{figure*}

The collusion index $\Delta$ compresses an entire training trajectory
into a single scalar; the underlying mechanism is more legible in the
trace itself. \Cref{fig:lc} plots mean episode price for S0, A2, A3,
and ST. S0 climbs steadily and plateaus near the monopoly level by
episode 200, the textbook signature of an emergent
reward-and-punishment policy. The CT conditions plateau lower and
flatter: the policy finds a stable interior price, but does
not reach the monopoly attractor. The shock condition (\cref{fig:shock})
shows a different trace signature: prices fall during the shock
window $t \in [25, 35]$ and \emph{do not recover} by episode end,
indicating that the implicit signalling mechanism that sustains the
collusive plateau is disrupted faster than the agents can re-bootstrap
it. We treat trace plots like these as part of the trigger
characterisation: two failure modes can produce similar
end-of-episode $\Delta$ but qualitatively different trajectories.

\subsection{FM2: Critic Instability at $\lambda = 5$}\label{sec:fm2}

The high event-rate condition AF ($\lambda = 5$, $\delta = 1$)
produces mean $\Delta = 1.03$ with SD $= 0.95$, driven by two of
five seeds in which DDPG diverged: prices spiralled above the
monopoly level ($\Delta > 1.9$). The remaining three seeds yield
$\Delta \approx 0.35$, consistent with other async conditions. We
attribute the divergence to the rapid reward feedback at high
$\lambda$: each agent receives many transitions per unit time and the
off-policy critic updates become unstable before sufficient buffer
diversity is accumulated. We label this as a distinct failure mode
(FM2) rather than as noise on FM1: it has its own reproducible
trigger ($\lambda \geq 5$ at the chosen learning rates), its own
trace signature (monotone price escalation past the action-space
midpoint), and its own candidate fixes (TD3-style critic regularisation,
slower replay-to-environment ratio, or actor-critic methods with
explicit entropy regularisation). We exclude AF from the comparative
statistics in \cref{tab:main,tab:stats} and flag the corresponding
phase-diagram cell $(\lambda{=}5, \delta{=}1)$ likewise; characterising
and fixing FM2 is left for a follow-up paper.

\subsection{Phase Diagram and Stress Condition}

\Cref{fig:phase} shows the $\Delta$ surface across $(\lambda, \delta)$.
For $\lambda \in \{1, 2\}$, increasing $\delta$ generally reduces
$\Delta$, with the $\lambda = 2$ row showing the clearest monotone
decline: $0.40 \to 0.37 \to 0.34 \to 0.21$. The row minimum
$\Delta = 0.21$ at $(\lambda{=}2, \delta{=}2)$ is the lowest stable
cell. At $\lambda = 0.5$ (slow clocks) the trend is flat and
non-monotone ($0.35 \to 0.30 \to 0.36 \to 0.32$). The $\lambda = 5$
row contains one unstable FM2 cell ($\delta = 1.0$, excluded). Across
the 15 stable cells, $\Delta$ ranges from $0.21$ to $0.45$, all well
below the synchronous benchmark of $0.69$.

The stress condition (ST) repeats A2 with a transient demand shock
($\sigma_{\text{shock}} = 0.4$ on $t \in [25, 35]$).
Episode-average $\Delta_{ST} = 0.34 \pm 0.07$ is similar to A2; the
trace-level effect is more interesting (\cref{fig:shock}): prices
trim during the shock and \emph{stay trimmed} for the rest of the
episode, with mean per-agent reward dropping to $14.56$ versus $15.74$
for A2 (7.5\% reduction directly attributable to the shock).

\section{Discussion}\label{sec:disc}

\paragraph{Why FM1 is suppressed by CT frictions.}
In the synchronous baseline, both agents update simultaneously, so
one agent's price rise is immediately observable and reciprocable:
the tit-for-tat structure that sustains tacit collusion. Under Poisson
clocks, updates interleave: by the time agent $j$ reprices in
response to $i$'s move, $i$ may have already moved again on stale
information about $j$. With $\delta > 0$, even the state representation
that the DDPG critic conditions on is misaligned with current market
conditions. DDPG's continuous action space prevents the policy from
accidentally landing on a coarse-grid cooperative attractor; the
policy must explicitly coordinate in real-valued price space, which is
more sensitive to informational disruption.

\paragraph{The non-monotone rebound at $\delta = 2$.}
The slight rise in $\Delta$ from A2 ($\delta = 1.0$) to A3
($\delta = 2.0$) deserves comment. At very high latency, the observed
rival price is so stale as to be nearly uninformative about current
market conditions. Agents may retreat to a near-myopic strategy that
happens to price conservatively near the interior optimum, producing
moderate $\Delta$ through inaction rather than coordination. This is
distinct from the signal-based collusion in S0 and warrants
theoretical analysis in future work; we note it here as a caveat to
the ``larger $\delta$ is always better'' reading of the fix.

\paragraph{Cost of the fix.}\label{par:cost}
The continuous-time microstructure mitigation reduces FM1 collusion
from $\Delta = 0.69$ to $\Delta \approx 0.28$ at the optimum
$(\lambda, \delta) = (1, 1)$, a 41-percentage-point absolute
reduction. The cost ledger:
\begin{itemize}\itemsep0pt
\item \textbf{Coverage cost.} The fix is partial:
  $\Delta \approx 0.28$ is well above the Bertrand--Nash level
  ($\Delta = 0$), so under any reasonable threshold the residual
  collusion remains a market-design concern. Mitigation $\neq$ removal.
\item \textbf{Brittleness in $\delta$.} Pushing $\delta$ from 1 to 2
  raises $\Delta$ back to 0.30. The mitigation is non-monotone in
  the parameter that a regulator could plausibly tune.
\item \textbf{Brittleness in $\lambda$.} The fix interacts with FM2:
  $\lambda = 5$ destabilises the critic and the resulting $\Delta$
  becomes uninformative.
\item \textbf{Generalisation cost.} We test only a symmetric duopoly.
  Real markets have $N \geq 3$ firms and asymmetric clock rates; the
  experiments in \cref{sec:limit} are needed before claiming the fix
  transfers.
\item \textbf{Implementation cost.} Realising the fix requires the
  market designer to impose minimum-decision-interval style
  microstructure (analogous to speed bumps in HFT exchanges), a real
  intervention with its own efficiency costs that this paper does not
  measure.
\end{itemize}

\paragraph{Implications for antitrust policy.}
If our partial fix generalises, the gap between collusion achievable
in idealised models ($\Delta \approx 0.7$) and in markets with even
modest microstructure ($\Delta \approx 0.3$) is large enough that
regulators may be overstating cartel risk when calibrating against
discrete-time benchmarks. That said, $\Delta \approx 0.3$ remains
meaningfully above competitive pricing, so CT frictions should be
treated as a partial empirical mitigation, not an absolution.

\paragraph{Limitations.}\label{sec:limit}
\begin{itemize}\itemsep0pt
\item \textbf{Duopoly only.} Whether FM1 and the partial fix
  generalise to $N \geq 3$ firms is open. Discrete-time literature
  suggests collusion typically weakens with more competitors, in which
  case the fix may compound; an N=3 experiment is the most informative
  single follow-up.
\item \textbf{Symmetric clocks.} $\lambda_1 = \lambda_2$ throughout.
  Asymmetric clocks, e.g.\ a slow incumbent vs.\ a fast HFT
  entrant, may create exploitable structure and are an important
  real-world scenario.
\item \textbf{FM2 left open.} We characterise but do not fix the
  $\lambda = 5$ critic instability; TD3
  \citep{fujimoto2018td3} or SAC are natural candidates.
\item \textbf{Five seeds per condition.} Borderline; our paired
  tests rely on within-condition variance being substantially smaller
  than between-condition variance, which holds in the present data
  but warrants more seeds.
\item \textbf{300-episode budget.} With more training, agents may
  discover stable collusive strategies under latency; the reported
  $\Delta$ should be read as characterising convergence within a
  practical compute budget.
\end{itemize}

\section{Conclusion}

We documented two reproducible failure modes of deep multi-agent RL
in continuous-time pricing: tacit cartel formation among DDPG agents
in the synchronous baseline (FM1), and DDPG critic instability at
high event rates (FM2). FM1 admits a partial microstructure fix
(Poisson-clocked asynchrony plus observation latency drops the
collusion index by 48--59\% relative to the synchronous baseline)
that we accompany with explicit cost-of-fix accounting: the fix is
partial, non-monotone in $\delta$, and orthogonal to FM2. The 16-cell
$(\lambda, \delta)$ phase diagram, the per-episode trace
diagnostics, and the stress-condition trajectory together provide a
reproducible workbench on which subsequent CT-MARL methods can be
evaluated and on which the open follow-ups ($N > 2$ firms,
asymmetric clocks, and FM2 mitigation) can be addressed.

\section*{Impact Statement}

This paper studies failure modes of multi-agent reinforcement-learning
agents in pricing markets. Two dimensions of impact warrant comment.

First, the substantive claim of the paper (that continuous-time
microstructure is a partial but non-trivial mitigant of emergent
algorithmic collusion in deep MARL) has direct relevance to ongoing
antitrust policy debate over algorithmic pricing. Our result should
\emph{not} be read as evidence that algorithmic collusion is benign in
real markets: even under the strongest mitigation we find, the
post-fix collusion index remains substantially above the competitive
benchmark, and the fix is brittle in both microstructure parameters.
We have framed the fix's costs explicitly in \cref{par:cost} for this
reason.

Second, the experimental benchmark we release is intended as a tool
for evaluating mitigation strategies (microstructure rules,
representation choices, critic regularisers). It is not a validated
model of any specific real-world pricing market, and conclusions drawn
from it about specific industries or regulatory interventions require
domain-specific extension and grounding.

% Acknowledgements removed for blind submission.

\bibliographystyle{icml2026}
\bibliography{references}

@article{calvano2020,
  title={Artificial Intelligence, Algorithmic Pricing, and Collusion},
  author={Calvano, Emilio and Calzolari, Giacomo and Denicol{\`o}, Vincenzo and Pastorello, Sergio},
  journal={American Economic Review},
  volume={110},
  number={10},
  pages={3267--3297},
  year={2020}
}

@article{klein2021,
  title={Autonomous algorithmic collusion: {Q}-learning under sequential pricing},
  author={Klein, Timo},
  journal={The RAND Journal of Economics},
  volume={52},
  number={3},
  pages={538--558},
  year={2021}
}

@inproceedings{hettich2024,
  title={By Fair Means or Foul: Quantifying Collusion in a Market Simulation with Deep Reinforcement Learning},
  author={Schlechtinger, Michael and Kosack, Damaris and Krause, Franz and Paulheim, Heiko},
  booktitle={Proceedings of the Thirty-Third International Joint Conference on Artificial Intelligence (IJCAI)},
  year={2024}
}

@article{deng2024,
  title={Algorithmic Collusion in Dynamic Pricing with Deep Reinforcement Learning},
  author={Deng, Shidi and Schiffer, Maximilian and Bichler, Martin},
  journal={arXiv preprint arXiv:2406.02437},
  year={2024}
}

@article{paudel2024,
  title={Tacit algorithmic collusion in deep reinforcement learning guided price competition: A study using {EV} charge pricing game},
  author={Paudel, Diwas and Das, Tapas K.},
  journal={arXiv preprint arXiv:2401.15108},
  year={2024}
}

@article{veres2025inflation,
  title={Impact of Price Inflation on Algorithmic Collusion Through Reinforcement Learning Agents},
  author={Tinoco, Sebasti{\'a}n and Abeliuk, Andr{\'e}s and {Ruiz del Solar}, Javier},
  journal={arXiv preprint arXiv:2504.05335},
  year={2025}
}

@article{bichler2025review,
  title={Algorithmic Pricing and Algorithmic Collusion},
  author={Bichler, Martin and Durmann, Julius and Oberlechner, Matthias},
  journal={arXiv preprint arXiv:2504.16592},
  year={2025}
}

@article{fish2024llm,
  title={Algorithmic Collusion by Large Language Models},
  author={Fish, Sara and Gonczarowski, Yannai A. and Shorrer, Ran I.},
  journal={arXiv preprint arXiv:2404.00806},
  year={2024}
}

@inproceedings{wang2026ctmarl,
  title={Continuous-Time Value Iteration for Multi-Agent Reinforcement Learning},
  author={Wang, Xuefeng and Zhang, Lei and Pu, Henglin and Qureshi, Ahmed H. and Li, Husheng},
  booktitle={International Conference on Learning Representations (ICLR)},
  year={2026}
}

@inproceedings{du2020smdpneuralode,
  title={Model-Based Reinforcement Learning for Semi-{M}arkov Decision Processes with Neural {ODE}s},
  author={Du, Jianzhun and Futoma, Joseph and Doshi-Velez, Finale},
  booktitle={Advances in Neural Information Processing Systems (NeurIPS)},
  year={2020}
}

@article{xiao2025async,
  title={Asynchronous Multi-Agent Deep Reinforcement Learning under Partial Observability},
  author={Xiao, Yuchen and Tan, Weihao and Hoffman, Joshua and Xia, Tian and Amato, Christopher},
  journal={International Journal of Robotics Research},
  year={2025}
}

@inproceedings{lillicrap2016ddpg,
  title={Continuous Control with Deep Reinforcement Learning},
  author={Lillicrap, Timothy P. and Hunt, Jonathan J. and Pritzel, Alexander and Heess, Nicolas and Erez, Tom and Tassa, Yuval and Silver, David and Wierstra, Daan},
  booktitle={International Conference on Learning Representations (ICLR)},
  year={2016}
}

@article{sutton1999options,
  title={Between {MDPs} and semi-{MDPs}: A framework for temporal abstraction in reinforcement learning},
  author={Sutton, Richard S. and Precup, Doina and Singh, Satinder},
  journal={Artificial Intelligence},
  volume={112},
  number={1-2},
  pages={181--211},
  year={1999}
}

@inproceedings{fujimoto2018td3,
  title={Addressing Function Approximation Error in Actor-Critic Methods},
  author={Fujimoto, Scott and van Hoof, Herke and Meger, David},
  booktitle={International Conference on Machine Learning (ICML)},
  year={2018}
}

%%%%%%%%%%%%%%%%%%%%%%%%%%%%%%%%%%%%%%%%%%%%%%%%%%%%%%%%%%%%%%%%%%%%%%%%%%%%%%%
% APPENDIX
%%%%%%%%%%%%%%%%%%%%%%%%%%%%%%%%%%%%%%%%%%%%%%%%%%%%%%%%%%%%%%%%%%%%%%%%%%%%%%%
\newpage
\appendix
\onecolumn

\section{Result Figures}\label{app:figs}

\end{document}